\documentclass[prapplied,reprint,superscriptaddress,amsmath,amssymb]{revtex4-2}

\usepackage{physics}
\usepackage{siunitx}
\AtBeginDocument{\RenewCommandCopy\qty\SI}
\usepackage{graphicx}
\usepackage{dcolumn}
\usepackage{bm}
\usepackage{comment}
\usepackage{booktabs}
\usepackage{amsmath}

\begin{document}
\title{Burst-Driven Collective Ultrafast Interactions for Confined Optical Absorption}

\author{Burak Arayıt}
\affiliation{Electrical and Electronics Engineering, Faculty of Engineering, Özyeğin University, Istanbul 34794, Türkiye}
\affiliation{ÖzU HealthTech Research Center, Özyeğin University, Çekmeköy, Istanbul, 34794, Türkiye}

\author{Seydi Yavaş}
\affiliation{ÖzU HealthTech Research Center, Özyeğin University, Çekmeköy, Istanbul, 34794, Türkiye}
\affiliation{Faculty of Engineering, Özyeğin University, Istanbul 34794, Türkiye}
\affiliation{Department of Physics, Boğaziçi University, 34342, Istanbul, Türkiye}
\affiliation{Lumos Laser A.Ş., 34340, Istanbul, Türkiye}

\author{Mehmet Burçin Ünlü}
\affiliation{ÖzU HealthTech Research Center, Özyeğin University, Çekmeköy, Istanbul, 34794, Türkiye}
\affiliation{Faculty of Engineering, Özyeğin University, Istanbul 34794, Türkiye}
\affiliation{Faculty of Aviation and Aeronautical Sciences, Özyeğin University, Istanbul 34794, Türkiye}

\author{Fatih Ömer Ilday}
\email{oemer.ilday@ruhr-uni-bochum.de}
\affiliation{Faculty of Electrical Engineering and Information Technology, Ruhr-Universität Bochum, 44801, Germany}
\affiliation{Faculty of Physics and Astronomy, Ruhr-Universität Bochum, 44801, Germany}
\affiliation{Center for Complex Interactions, Ruhr-Universität Bochum, 44801, Germany}

\begin{abstract}
Ultrafast bursts, groups of closely spaced femtosecond pulses, can drive light–matter interactions that are not possible with single-pulse excitations. This occurs when earlier pulses create a transient effect that nonlinearly alters the interaction of later pulses, provided that the inter-pulse spacing is shorter than the relaxation time of the transient. A long-standing trade-off in optical excitation is between multiphoton absorption, which provides strong spatial confinement but low energy-transfer efficiency, and linear absorption, which is efficient but offers limited spatial localization. Here, we show that a burst of ultrafast pulses can collectively build up a population of free carriers in semiconductor nanoparticles via multiphoton absorption. The free carriers absorb linearly and strongly, and dissipate rapidly once the burst ends. The result is a more efficient absorption process that is strongly gated spatially and temporally by multiphoton absorption and burst duration, respectively. We develop an analytical model that couples carrier generation with Auger and surface recombination, yielding a piecewise-exact description of the carrier dynamics and an intuitive scaling relation for the collective energy-deposition enhancement. A signature consequence of collective absorption is its dependence on one extra power of the light intensity, effectively raising the order of an $m$-photon process to $m+1$ without requiring higher intensities for the individual pulses. While we focus on nonlinear photoacoustic imaging as a potential application, we anticipate diverse applications, including nonlinear fluorescence microscopy.
\end{abstract}

\maketitle

\section{Introduction}

Nonlinear optical absorption confines the interaction volume to the focal region, providing optical sectioning and selectivity that affords multiphoton microscopy and related imaging techniques a decisive advantage \cite{Helmchen-2005, wang2018three}. However, this spatial confinement comes at the cost of low energy-transfer efficiency, as the nonlinear excitation cross-sections are much smaller than their linear counterparts, thereby requiring ultrashort pulses with high peak intensities \cite{xu1996multiphoton}. These intensities, and thus the achievable efficiencies, cannot be increased indefinitely without exceeding damage thresholds \cite{hopt-2001, koenig2000multiphoton}. In contrast, linear absorption enables efficient energy deposition but lacks spatial selectivity, leading to signal generation outside the focal volume, especially axially \cite{liu2012effects,theer-2006}. This trade-off between localization and efficiency is regarded as a fundamental limitation of optical excitation schemes.

Ultrafast burst-mode excitation provides a route to overcome this limitation by distributing the energy over a large number of pulses of moderate energy within an excitation time window. However, there is a further possibility, namely, to exploit what we call collective pulse–matter interactions, accessed via ultrafast bursts \cite{kerse2016}. If the inter-pulse spacing is much shorter than the characteristic relaxation time \cite{radu2011transient, krishnamoorthy2019optical}, transient material responses to early pulses persist and accumulate toward a nonequilibrium state. Burst-driven collective behavior has been demonstrated in ablation-cooled material removal~\cite{kerse2016}, enabling more efficient ablation~\cite{Sugioka2021}.

\begin{figure*}[t]
    \centering
    \includegraphics[width=\textwidth]{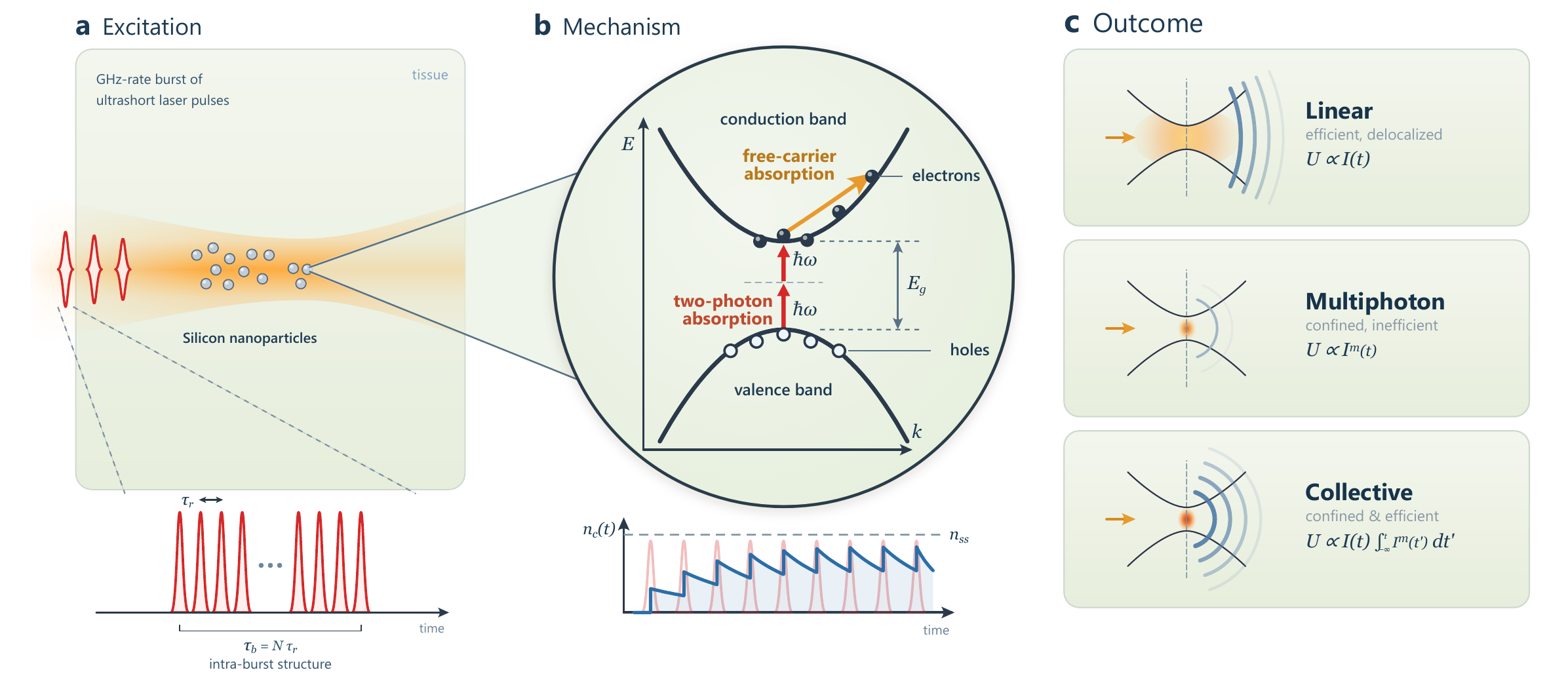}
    \caption{Schematic of burst-driven collective absorption. (a)~A GHz-rate burst of ultrashort pulses is focused into a cluster of silicon nanoparticles. The inset shows the intra-burst pulse train, with inter-pulse period $\tau_\text{r}$ and burst duration $\tau_\text{b}=N\tau_\text{r}$. (b)~Per-pulse mechanism. Two-photon absorption (TPA) generates free carriers that subsequently absorb through free-carrier absorption (FCA), so the carrier density $n_\text{c}(t)$ builds up from pulse to pulse toward a steady state $n_\text{ss}$. Each jump in $n_\text{c}(t)$ marks one pulse of the burst. (c)~Resulting excitation regimes, with the corresponding absorbed-energy scaling (common absorption coefficients omitted). The linear regime [$U\propto I(t)$] is efficient but delocalized, the multiphoton regime [$U\propto I^m(t)$] is confined but inefficient, and the collective regime [$U\propto I(t)\int_{-\infty}^{t} I^m(t')\,dt'$] is confined and efficient.}
    \label{fig:schematic}
\end{figure*}

Here, we introduce a collective nonlinear absorption mechanism, enabled by ultrafast bursts, where nonlinear absorption can act as a spatially confined trigger that opens a finite-time window of effectively linear and efficient absorption. We focus on multiphoton carrier injection in semiconductor nanoparticles. At low-picosecond inter-pulse separations, pulses arrive much faster than carrier recombination~\cite{driel1986,othonos1998}. Each pulse generates carriers through nonlinear absorption; subsequent pulses interact with the remaining carriers, experience enhanced linear absorption, and generate still more carriers, closing the feedback loop. After sufficient buildup, the interaction transitions from being merely seeded by nonlinear excitation to being dominated by carrier-assisted linear absorption. Crucially, this efficient absorption occurs only within the spatial region where the carriers have been generated via multiphoton absorption, thereby retaining the spatial confinement of the nonlinear excitation. Our key finding can be summarized as a transformation from single-pulse nonlinear excitation to collective nonlinear excitation  [Fig.~\ref{fig:schematic}]:
\begin{equation}
 U\propto I^m(t) \; \rightarrow  \; I(t)\int_{-\infty}^{t} I^m(t')\,dt'.
 \label{nonlinear_to_collective_nonlinear}
\end{equation}

This transformation circumvents the limitation that the term, $ I^m(t)$, is strongly capped to avoid photodisruption: its integral over many pulses can be larger by at least the number of pulses per burst, and potentially even larger through cooperative effects, as we demonstrate below. Equally important, the right-hand side contains $m+1$ factors of the intensity. Collectivity therefore converts an $m$-photon excitation into an effectively $(m+1)$-photon one, raising the order of the nonlinearity by one without the vanishingly small cross-section of a true $(m+1)$-photon transition, and sharpening the spatial confinement accordingly. This way, we combine efficient energy deposition without sacrificing spatial confinement, effectively combining the advantages of nonlinear and linear excitation [Fig.~\ref{fig:schematic}]. Once the burst ends, the accumulated carriers quickly recombine and this window closes on the recombination timescale, rendering the enhancement intrinsically time-gated.

As one prominent application, we focus on addressing a long-standing problem in photoacoustic signal generation, which directly depends on localized energy deposition \cite{wang-2007}. Multiphoton photoacoustics offers greater contrast but suffers from weak nonlinear absorption and poor signal-to-noise ratio, which has limited its practical applicability \cite{yamaoka2014photoacoustic-827,lai2014nonlinear,langer2013twophoton-3d6}. In contrast, linear excitation offers poor spatial confinement, and out-of-focus background absorption progressively degrades image contrast with increasing imaging depth \cite{liu2012effects}. Ultrafast burst-driven collective absorption could address this limitation by enhancing the nonlinear photoacoustic signal by two to three orders of magnitude over single-pulse excitation while preserving the focal confinement of multiphoton excitation within a sufficiently short time window, set by the burst duration, that satisfies stress confinement conditions. Semiconductor nanoparticles, widely used in biomedical applications \cite{ofarrell2006,peng2014}, provide a particularly suitable platform for this mechanism owing to their well-characterized multiphoton absorption and free-carrier response \cite{boggess1986}. When a semiconductor is targeted, multiphoton absorption can generate free charge carriers \cite{driel1986}, thereby introducing an additional linear absorption channel for subsequent pulses. By using temporally closely spaced pulses, carrier recombination can be minimized while cumulative energy delivery is increased \cite{othonos1998}.

We first develop an analytical model describing burst-driven collective absorption mediated by free-carrier dynamics in semiconductors. By coupling multiphoton carrier generation with Auger and surface recombination, we obtain a pulse-resolved description of carrier buildup and derive closed-form expressions for the resulting enhancement in energy deposition. The model predicts a transition from nonlinear excitation to carrier-assisted absorption and provides a framework for understanding confined and efficient energy deposition under ultrafast collective excitation. We show that confinement is preserved in the presence of scattering. We then focus on photoacoustic imaging.

\section{Theory}

\subsection{Absorption and energy coupling}
\label{sec:absorption}

Let  $U(t, x, y, z)$ be the absorbed energy density and $\partial U/\partial t$ its local rate of change. Including intrinsic linear absorption, $\mu_\text{a}^\text{int}$, multi-photon absorption of order $m$, with coefficient $\beta_m$, and linear absorption by free carriers, we write
\begin{equation}
    \frac{\partial U}{\partial t} = \left(\mu_\text{a}^\text{int} + \sigma\, n_\text{c}(t)\right) I(t) + \beta_m\,I^m(t),
    \label{eq:heating_continuous}
\end{equation}
where $n_\text{c}(t)$ is the free-carrier density, governed by Eq.~\eqref{eq:n_fc} below, and $\sigma$ is the free-carrier absorption cross section~\cite{sokolowskitinten2000, driel1986}. The free-carrier population density, $n_\text{c}$, is typically extremely low at room temperature under equilibrium conditions for most materials. Consequently, there are two conventional cases, plus the collective interaction introduced here: (i) At wavelengths where the photon energy is above the band gap, linear absorption is typically the dominant mechanism. Energy coupling extends along the entire beam path [top panel of Fig.~\ref{fig:schematic}(c)], and spatial localization is weak, particularly along the beam axis. Any free-carrier generation due to linear absorption effectively increases the strength of absorption but does not change this picture qualitatively.  (ii) At wavelengths below the band gap, linear absorption vanishes, and multiphoton absorption becomes the primary mechanism of coupling of ultrafast pulses. Its nonlinear dependence on intensity confines the interaction to the focal volume [middle panel of Fig.~\ref{fig:schematic}(c)], at the cost of low efficiency. These two limits define the standard localization--efficiency trade-off. As we demonstrate below, this trade-off can be overcome by building a non-equilibrium free carrier population throughout a burst of ultrafast pulses [Fig.~\ref{fig:schematic}(a), (b), and bottom panel of (c)]. (iii) Under burst excitation, each pulse is shorter than all other timescales of the interaction and therefore samples the carrier density at its pre-pulse value $n_{\text{c},k}$, so Eq.~\eqref{eq:heating_continuous} reduces to the per-pulse expression
\begin{equation}
    \frac{\partial U_k}{\partial t} = (\mu_\text{a}^\text{int} + \sigma n_{\text{c},k})\,I(t) + \beta_m\,I^m(t),
    \label{eq:heating_modified}
\end{equation}
the discrete expression of the history-dependent absorption depicted in Fig.~\ref{fig:schematic}(b). The $\sigma n_{\text{c},k}$ contribution is a linear absorption channel that did not exist before the burst began, created collectively by the preceding pulses through carrier accumulation. Because the carrier population that enables this channel was seeded by multiphoton absorption, it is spatially confined to the focal volume. The system therefore achieves efficient energy deposition without sacrificing spatial selectivity, circumventing the fundamental trade-off identified in the introduction.

The energy absorbed through Eq.~\eqref{eq:heating_modified} ultimately becomes heat, along two routes. A multiphoton absorption event delivers the combined photon energy $m\hbar\omega$. The excess $m\hbar\omega - E_\text{g}$, where $E_\text{g}$ is the band-gap energy, appears as kinetic energy of the generated electron--hole pair and is transferred to the lattice through carrier--phonon coupling within a few picoseconds, while the remaining $E_\text{g}$ is stored in the pair and is released through the same channel upon nonradiative recombination~\cite{driel1986, othonos1998}. Free-carrier absorption instead transfers $\hbar\omega$ directly to the kinetic energy of an existing carrier without creating a pair, and thus thermalizes promptly~\cite{driel1986, sokolowskitinten2000}. Since radiative recombination is negligible for the indirect-gap materials considered here, essentially all of the deposited energy converts to lattice heat. Carrier generation stores part of the absorbed energy, and recombination subsequently releases it.

We now model the evolution of $n_\text{c}$ under burst excitation.

\subsection{Carrier dynamics under burst excitation}
\label{sec:carrier_dynamics}

The temporal separation of the pulses within the burst is defined by $\tau_\text{r}$, and the duration of the burst is $\tau_\text{b}$, corresponding to $N=\tau_\text{b}/\tau_\text{r}$ pulses within the burst.

A photon of sufficient energy can promote an electron from the valence band to the conduction band. High-intensity subpicosecond pulses at longer wavelengths drive this process via multiphoton absorption, where $m\hbar\omega > E_\text{g}$ and $m$ is the dominant order of multiphoton absorption. The carrier generation rate is governed by
\begin{equation}
\frac{dn_\text{c}}{dt} = \frac{\mu_\text{a}^\text{int}}{\hbar\omega}\,I(t) + \frac{\beta_m}{m\hbar\omega}\,I^{m}(t) - \gamma\,n_\text{c}^3 - \frac{n_\text{c}}{\tau_\text{eff}},
\label{eq:n_fc}
\end{equation}
where $n_\text{c}$ is the free-carrier density, $\mu_\text{a}^\text{int}$ is the intrinsic linear absorption coefficient, $\beta_m$ is the $m$-photon absorption coefficient, $\gamma$ is the Auger recombination constant \cite{othonos1998,driel1986}, and $\tau_\text{eff}$ is a generalized effective linear carrier loss time. The first two terms describe carrier generation via linear and multiphoton absorption, respectively. However, our focus is on material and wavelength combinations with negligible linear absorption, as in the sub-gap excitation of crystalline silicon developed below. Therefore, $\mu_\text{a}^\text{int}\approx0$, and carrier injection is seeded almost entirely by multiphoton absorption. This distinction is important physically: intrinsic linear absorption would generate carriers wherever the beam deposits energy, whereas the collective absorption considered here is initiated by nonlinear absorption and is therefore confined to the focal region. The third term accounts for Auger recombination, in which an electron-hole pair recombines, transferring its energy to a third carrier. This process depletes the carrier population with a third-order dependence on $n_\text{c}$. Other carrier loss channels, such as ambipolar and surface diffusion together with Shockley--Read--Hall recombination, are represented by a lumped first-order loss term with an effective decay time, $\tau_\text{eff}$. While radiative recombination is also relevant for direct-gap semiconductors, Auger remains the dominant recombination channel for silicon \cite{othonos1998}.

\begin{figure*}[tp]
    \centering
    \includegraphics[width=\textwidth]{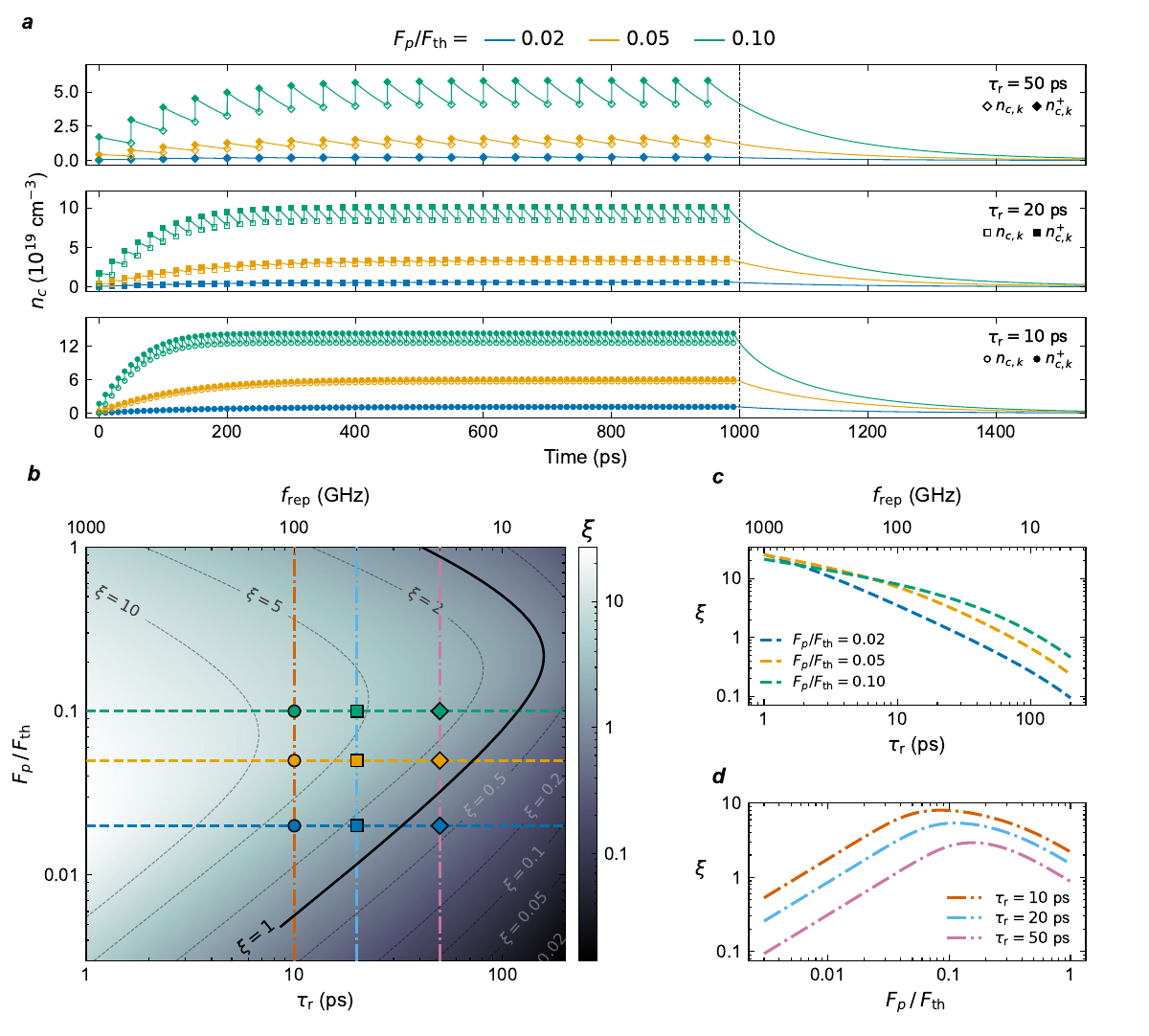}
    \caption{Carrier dynamics and the collective regime for c-Si nanoparticles. (a) Free-carrier density evolution exhibiting the sawtooth characteristic during the burst at inter-pulse periods $\tau_\text{r} = 10$, $20$, and $50$~\si{ps}, each at per-pulse fluences $F_\text{p} = 0.02$, $0.05$, and $0.1\,F_\text{th}$. Open and filled markers denote the pre- and post-pulse densities $n_{\text{c},k}$ and $n_{\text{c},k}^{+}$, with a distinct marker shape per inter-pulse period (circle, square, diamond). (b) Collective parameter $\xi \equiv \sigma n_\text{ss}/(\beta_m I_\text{p}^{m-1})$, the ratio of the free-carrier absorption channel to the direct multiphoton absorption channel, across the ($\tau_\text{r}$, $F_\text{p}/F_\text{th}$) parameter space. The top axis gives the corresponding intra-burst repetition rate $f_\text{rep} = 1/\tau_\text{r}$. Thin contours mark $\xi$ at the standard 1--2--5 levels per decade with each level labeled, and the solid $\xi = 1$ contour marks where the two heating channels contribute equally. Dashed horizontal and dash-dotted vertical lines mark the fluence and inter-pulse-period cuts profiled in (c) and (d), with markers at their crossings keyed to the nine scenarios of (a). (c) $\xi$ as a function of $\tau_\text{r}$ at the three fluences of (a). The dashed line style matches the horizontal cuts in (b). (d) $\xi$ as a function of $F_\text{p}/F_\text{th}$ at the three inter-pulse periods of (a). The dash-dotted line style matches the vertical cuts in (b).}
    \label{fig:dynamics}
\end{figure*}

To solve Eq.~\ref{eq:n_fc}, we separate the dynamics into two phases. During the ultrashort pulse, only generation occurs, and recombination is negligible because of the separation of timescales, $\tau_\text{p} \ll \tau_\text{r} < \tau_\text{eff}$, where $\tau_\text{p}$ is the pulse width, and $\tau_\text{r}$ is the time between the pulses. Between pulses, recombination reduces the carrier density via the mechanisms described above. This separation yields a pair of coupled difference equations, where $n_{\text{c},k}^+$ denotes the carrier density immediately after pulse $k$ and $n_{\text{c},k+1}$ is the carrier density just before the next pulse,
\begin{subequations}
\label{eq:recurrence}
\begin{align}
n_{\text{c},k}^+ &= n_{\text{c},k}  + \frac{\beta_m\,I_\text{p}^m\,\tau_\text{p}}{m\hbar\omega}, \label{eq:recurrence_a} \\[6pt]
n_{\text{c},k+1} &= \frac{n_{\text{c},k}^+\,e^{-\tau_\text{r}/\tau_\text{eff}}}{\sqrt{1 + \gamma\tau_\text{eff}\,(n_{\text{c},k}^+)^2\!\left(1 - e^{-2\tau_\text{r}/\tau_\text{eff}}\right)}}, \label{eq:recurrence_b}
\end{align}
\end{subequations}
with $n_{\text{c},0} \approx 0$ and $I_\text{p}$ the peak pulse intensity.  Eq.~\ref{eq:recurrence_a} represents the injection of carriers through nonlinear absorption. Eq.~\ref{eq:recurrence_b} describes the nonlinear recombination that reduces the carrier density over the inter-pulse period $\tau_\text{r}$. When $\tau_\text{r}$ is sufficiently short, recombination cannot deplete all the carriers produced and the material remains in its transient state when the next pulse arrives. The system is therefore driven repeatedly away from equilibrium, with the carrier density building up pulse by pulse until it approaches a steady state $n_\text{ss}$, defined by $n_{\text{c},k+1} = n_{\text{c},k} \equiv n_\text{ss}$, where per-pulse injection is exactly balanced by inter-pulse recombination. This results in a sawtooth-like fast pattern with an initially rising, then plateauing baseline [Fig.~\ref{fig:dynamics}(a)]. The carrier density evolves exactly as the surface temperature does in ablation-cooled material removal, where each pulse of a GHz burst deposits heat faster than it can dissipate between pulses~\cite{kerse2016}. In both cases, each pulse drives the relevant state variable away from equilibrium, free carriers here and residual heat in ablation cooling, thereby strengthening the action of later pulses until a competing loss mechanism, Auger recombination here and ablation itself in the latter, limits the growth.

The sustained transient non-equilibrium state persists throughout the burst duration, during which the material's optical and thermal characteristics also become transient. Carriers are promoted to the conduction band very rapidly, and their kinetic energy simultaneously rises under the intense field of the incoming pulse, driving the electron temperature $T_\text{e}$ \cite{driel1986, othonos1998}.

With the elevated carrier temperature, the impact ionization rate also increases, as energetic carriers generate additional electron--hole pairs and reinforce the positive feedback beyond multiphoton-induced carrier generation~\cite{driel1986}. Moreover, the excess carrier concentration reduces the bandgap, causing bandgap renormalization (BGR), with $\Delta E_\text{g}(n_\text{c})\propto n_\text{c}^{1/3}$~\cite{driel1986, sokolowskitinten2000}, further accelerating carrier generation. These effects constitute additional positive feedback channels that would lower the transition threshold and elevate the steady-state carrier density beyond the values predicted here. As the carrier density gets close to the critical carrier density for dielectric breakdown, $n_\text{c} \geq n_\text{cr} \sim 10^{21}$~\si{cm^{-3}}, BGR and impact ionization effects are non-negligible; they can cause runaway effects and might lead to avalanche breakdown~\cite{pronko1998}. However, we assume $n_\text{c} \leq n_\text{cr}$, and the present model omits both, retaining only multiphoton generation and Auger recombination as the essential competing processes.

\subsection{Collective energy deposition}
\label{sec:collective_energy}

With $n_{\text{c},k}$ given by the recurrence relations~\eqref{eq:recurrence}, we evaluate the total energy deposited during the burst. This outcome is driven by a nonlinear feedback loop in which each pulse generates carriers via multiphoton absorption, thereby enhancing the linear absorption of subsequent pulses, which in turn deposit energy more efficiently into the existing carrier population. More generally, collective light–matter interactions can arise through nonlinear feedback whenever a material state created by earlier excitation persists long enough to modify subsequent interactions~\cite{oektem2013nonlinear,Tokel2017}. Ultrafast bursts extend this concept to highly transient material states by delivering subsequent pulses before that state relaxes. The early pulses in the burst serve as sacrificial excitation~\cite{kerse2016}, investing energy inefficiently to build the carrier population and enabling later pulses to benefit from the collectively established absorption channel. Auger recombination provides a competing negative feedback that grows with carrier density and ultimately clamps the buildup at the steady state $n_\text{ss}$.

Integrating Eq.~\eqref{eq:heating_modified} over each pulse and summing over all $N$ pulses yields the total deposited energy density
\begin{equation}
U_\text{burst} = N\left(\mu_\text{a}^\text{int} + \beta_m\, I^{m-1}_\text{p}\right) I_\text{p}\tau_\text{p} + \sigma\, I_\text{p}\tau_\text{p} \sum_{k=1}^{N} n_{\text{c},k},
\label{eq:U_burst}
\end{equation}
where $I_\text{p}$ is the peak pulse intensity and $\tau_\text{p}$ is the pulse duration, assuming a square temporal profile. The summation $\sum n_{\text{c},k}$ represents the cumulative contribution of free-carrier absorption across all pulses, the collective energy-deposition term that has no counterpart in single-pulse excitation.

We next derive the scaling relation for the collective deposition regime shown schematically in Fig.~\ref{fig:schematic}(c). We first write the carrier density before the $k$-th pulse as the accumulated response to all preceding pulses. Let $I_j$ denote the peak intensity of the $j$-th pulse with the carrier density injected by that pulse in the absence of recombination during the pulse given by
\begin{equation}
\Delta n_j = \frac{\beta_m I_j^m \tau_\text{p}}{m\hbar\omega}.
\end{equation}
We can account for the inter-pulse carrier recombination through a memory kernel $\mathcal{K}_{k-j}$, with $\mathcal{K}_0=1$. Then, the pre-pulse carrier density can be written to leading order as
\begin{equation}
n_{\text{c},k} \simeq \sum_{j<k} \mathcal{K}_{k-j} \Delta n_j =
\frac{\beta_m \tau_{\rm p}}{m\hbar\omega} \sum_{j<k} \mathcal{K}_{k-j} I_j^m,
\label{eq:nc_memory_discrete}
\end{equation}
where $\mathcal{K}_{k-j}=\exp[-(k-j)\tau_\text{r}/\tau_\text{eff}]$. We focus on the limit of large number of pulses per burst, $N \gg 1$, and we introduce a continuous pulse coordinate $\nu\equiv(k-1)/(N-1)$ that ranges from 0 to 1. Replacing the sequence of peak intensity of the individual pulses, $I_j$, by $I(\nu)$ with $t = \nu \, \tau_{\rm b}$, where the burst duration is $\tau_{\rm b} = N\,\tau_{\rm r}$, Eq.~\eqref{eq:nc_memory_discrete} becomes 
\begin{equation}
n_\text{c}(\nu) \simeq \frac{\beta_m \tau_\text{p}}{m\hbar\omega \Delta \nu}
\int_0^\nu \mathcal{K}\!\left((\nu-\upsilon)\tau_\text{b}\right)\,I^m(\upsilon)\,d\upsilon,
\label{eq:nc_memory_continuous_u}
\end{equation}  
where $\Delta\nu \simeq \tau_\text{r}/\tau_\text{b}$. Changing variable to physical time, we get
\begin{equation}
n_\text{c}(t) \simeq \frac{\beta_m \tau_\text{p}}{m\,\hbar\omega\,\tau_\text{r}} \int_{-\infty}^{t} \mathcal{K}(t-t')\,I^m(t')\,dt'.
\label{eq:nc_memory_continuous_t}
\end{equation}
Substituting Eq.~\eqref{eq:nc_memory_continuous_t} into the free-carrier absorption term of Eq.~\eqref{eq:heating_modified} gives the collective contribution to the energy deposition,
\begin{eqnarray}
\dot U
&=& \left(\mu_\text{a}^\text{int} +  \frac{\sigma\beta_m \tau_\text{p}}{m\hbar\omega\,\tau_\text{r}} \int_{-\infty}^{t} \mathcal{K}(t-t')\,I^m(t')\,dt' \right) I(t) \nonumber\\ 
&& +\, \beta_m\,I^m(t).
\label{eq:collective_memory_heating}
\end{eqnarray}
The regime of interest here is when the collectively induced free-carrier absorption dominates over both the intrinsic linear absorption and the instantaneous multiphoton absorption of the current pulse. At sufficiently high intra-burst repetition rates, carrier losses between successive pulses become small and $\mathcal{K}(t-t')\simeq1$. The collective energy-deposition term then reduces to
\begin{equation}
\dot U \propto I(t) \int_{-\infty}^{t} I^m(t')\,dt'.
\label{eq:collective_scaling}
\end{equation}
As anticipated in Eq.~\eqref{nonlinear_to_collective_nonlinear}, the collective term carries $m+1$ powers of the intensity, so the deposition inherits the spatial profile of an $(m+1)$-photon process. This is why the collective deposition in Fig.~\ref{fig:confinement} is more tightly confined than the multiphoton deposition that seeds it, with axial widths of 11 and 13~\si{\micro m}, respectively.

\subsection{Energy-deposition enhancement}
\label{sec:collective_onset}

The collective buildup determines how efficiently the burst deposits energy. With the steady state established within the first pulses of the burst, $n_{\text{c},k} \approx n_\text{ss}$, and for negligible intrinsic absorption ($\mu_\text{a}^\text{int}\approx0$) the deposited energy density of Eq.~\ref{eq:U_burst} becomes
\begin{equation}
    U_\text{burst} \approx N\left(\beta_m I^{m-1}_\text{p} + \sigma n_\text{ss}\right) I_\text{p}\tau_\text{p}.
    \label{eq:U_burst_ss}
\end{equation}
Normalizing by the incident burst fluence, $F_\text{burst} = N I_\text{p}\tau_\text{p} = N F_\text{p}$, defines the energy-deposition coefficient of the burst,
\begin{equation}
    \eta \equiv \frac{U_\text{burst}}{F_\text{burst}} \approx \beta_m I^{m-1}_\text{p} + \sigma n_\text{ss},
    \label{eq:efficiency_definition}
\end{equation}
the sum of the multiphoton and free-carrier absorption coefficients. For pure $m$-photon excitation the same quantity is $\eta_\text{MPA} = \beta_m I^{m-1}_\text{p}$, so
\begin{equation}
    \frac{\eta}{\eta_\text{MPA}}\approx 1+\xi, \qquad \xi \equiv \frac{\sigma \,n_\text{ss}}{\beta_m \,I^{m-1}_\text{p}},
    \label{eq:efficiency_ratio}
\end{equation}
where $\xi$ is the ratio of FCA to MPA energy deposition at steady state. For $\xi \ll 1$ each pulse acts independently under multiphoton absorption. For $\xi > 1$, the collectively sustained carrier population becomes the dominant absorption channel.

\subsection{Realization in silicon nanoparticles}

To evaluate the predictions of the model, we apply it to crystalline silicon nanoparticles as a representative nonlinear absorber.

For finite absorbers such as these nanoparticles, carrier transport must be interpreted differently from bulk diffusion. In a bulk material, diffusion can redistribute carriers over distances that match or exceed the focal spot size, $\omega_0$, and thereby modify the spatial profile of absorption. In a nanoparticle with diameter $d \ll \omega_0$, however, carriers remain confined to the particle volume, and diffusion within the nanoparticle cannot transport carriers out of the focal volume, nor does it appreciably change the local optical intensity sampled by the carrier population. Its primary consequence is instead to bring carriers to the nanoparticle surface, where surface recombination provides an additional loss channel.

We account for this finite-size effect through an effective first-order carrier lifetime. Surface recombination removes carriers at a rate set by the surface recombination velocity $S$ acting over the particle's surface-to-volume ratio $A/V$. For a sphere of diameter $d$ this ratio is $A/V = 6/d$, so the surface-recombination-limited lifetime is
\begin{equation}
    \frac{1}{\tau_{\rm s}} = S\,\frac{A}{V} = \frac{6S}{d}, \qquad \tau_{\rm s} \simeq \frac{d}{6S}.
\end{equation}
For a silicon nanosphere with $d=100$ nm and $S=10^4$ cm/s~\cite{grumstrup2014ultrafast-7ad}, this gives $\tau_\text{s}\approx 170$ ps. This timescale is comparable to the upper limit of the inter-pulse periods considered here and can reduce the carrier density between pulses. In our model, we include this contribution by taking $\tau_\text{eff}$ to represent the relevant first-order carrier-loss time, dominated by surface recombination for the nanoparticle parameters used below.

As long as the particle diameter satisfies $d \gg a_\text{B}$, where $a_\text{B}$ is the Bohr exciton radius, quantum-confinement effects are negligible and the bulk values of $\beta_m$, $\gamma$, and $\sigma$ can be used as a first approximation~\cite{brus1984, delerue1993theoretical-46d}. We specialize to two-photon absorption ($m=2$) and select an excitation wavelength of $\lambda=1700$~\si{nm}, offering high penetration depth in biological tissue~\cite{kobat-2009, horton2013}. At this wavelength, silicon has negligible linear absorption ($\mu_\text{a}^\text{int}\approx 0$), so the TPA and FCA channels dominate. We take $\beta_m$ from Bristow et al.~\cite{bristow2007} and $\sigma$ from Boggess et al.~\cite{boggess1986} scaled by $\lambda^2$ to account for the free-carrier Drude response. The excitation parameters and material properties are presented in Table~\ref{tab:params}. The damage (ablation) threshold fluence for silicon is $F_\text{th}=0.2$~\si{J/cm^2} for ultrashort pulses \cite{bonse2002femtosecond-b0a}.

Figure~\ref{fig:dynamics}(a) shows how $n_\text{c}$ evolves over the burst duration, pulse by pulse, displaying a characteristic sawtooth pattern resulting from Eqs.~\ref{eq:recurrence}. Each pulse produces a sharp carrier increment, followed by nonlinear decay between pulses. This interplay continues until the carrier density reaches a steady state, $n_\text{ss} = n_{\text{c},k} = n_{\text{c},k+1}$, at which per-pulse generation is exactly balanced by inter-pulse recombination. The time to reach steady state decreases with increasing fluence, as stronger TPA generation drives the system toward equilibrium faster. Shorter $\tau_\text{r}$ allows less inter-pulse decay, leading to both higher $n_\text{ss}$ and faster convergence. Although the fluence values are chosen below the reported damage threshold of silicon, the critical carrier density $n_\text{cr}$ remains an important consideration.

Figure~\ref{fig:dynamics}(b) maps $\xi$ (Eq.~\ref{eq:efficiency_ratio}) across the ($\tau_\text{r}$, $F_\text{p}/F_\text{th}$) parameter space, delineating the boundary between the TPA-dominated ($\xi < 1$) and collective ($\xi > 1$) regimes. Panel~(c) shows the $\tau_\text{r}$ dependence at the three per-pulse fluences of panel~(a). $\xi$ rises steeply as $\tau_\text{r}$ decreases below the surface recombination time $\tau_\text{s} \approx 170$\,ps, because incomplete inter-pulse recovery allows $n_\text{ss}$ to accumulate to levels where FCA dominates over TPA. Panel~(d) shows the fluence dependence at the three inter-pulse periods of panel~(a). $\xi$ grows with $F_\text{p}$ at sub-threshold fluences as stronger TPA generation raises $n_\text{ss}$. At higher fluences, Auger recombination (third order in $n_\text{c}$) clamps $n_\text{ss}$ and slows the growth of $\xi$. Together, the two panels show that GHz intra-burst repetition rates and sub-threshold per-pulse fluences are the key conditions for accessing the collective regime.

\begin{figure*}[t]
    \centering
    \includegraphics[width=\textwidth]{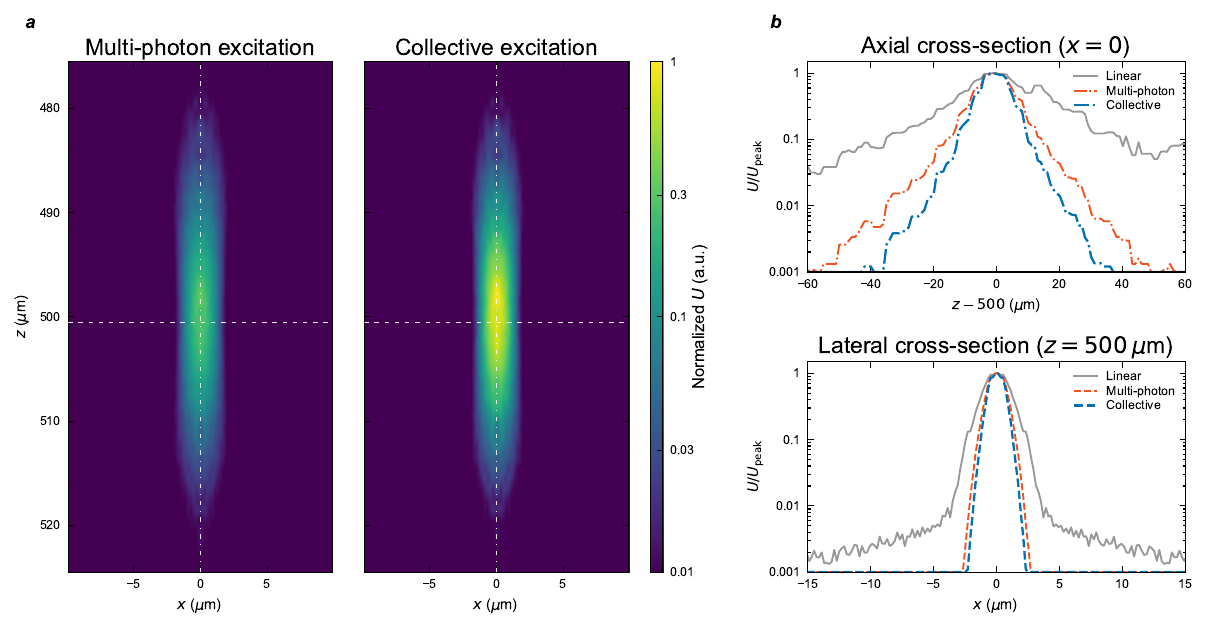}
    \caption{Confined energy deposition in brain tissue at 1700~\si{nm} under Monte Carlo light transport. (a) Cross-sectional absorbed energy density $U$ for multiphoton excitation and collective excitation, on a shared normalized scale (both divided by the same global peak, so the collective-over-multiphoton magnitude difference is preserved). The collective burst makes the focal nonlinear (TPA\,+\,FCA) signal both stronger and more tightly confined, with the focal volume showing its characteristic elongation along the optical axis. Dash-dotted and dashed guide lines mark the on-axis (axial) and focal-plane (lateral) cuts profiled in (b). (b)~Axial and lateral profiles, each normalized to its own peak, with color denoting the regime and line style the cut direction (axial dash-dotted, lateral dashed) to match the guides in (a). The linear absorption background is drawn as a solid gray reference in both profiles. The collective deposition is the most confined, with axial FWHMs of $11$, $13$, and $27$~\si{\micro m} and lateral FWHMs of $1.5$, $1.7$, and $2.4$~\si{\micro m} for the collective, multiphoton, and linear regimes, respectively. $\tau_\text{r}=10$~\si{ps}, $N=100$, $F_\text{p}=0.05\,F_\text{th}$.}
    \label{fig:confinement}
\end{figure*}

Figure~\ref{fig:confinement} shows the focal energy deposition across the excitation regimes under Monte Carlo light transport using parameters representative of brain tissue. The absorbed energy density $U(\mathbf{r})$ follows from Eq.~\eqref{eq:heating_modified}, and the fluence maps $\Phi(x,z)$ are obtained from Monte Carlo light transport~\cite{wang1995mcml} with a hyperboloid-focusing launch that reproduces the diffraction-limited focal waist~\cite{liu2012effects} for a focused Gaussian beam in brain at 1700~\si{nm} (Table~\ref{tab:params})~\cite{horton2013,Jacques2013}. The TPA and collectively gated FCA contributions are driven only by the ballistic, unscattered fluence, which retains the femtosecond Gaussian pulse profile; multiply scattered light is temporally dispersed and feeds the linear channel alone, its contribution to the focal nonlinear signal being negligible~\cite{theer-2006}. Linear absorption produces a broad, delocalized background along the beam path, a single fs pulse (MPA) confines deposition to the focal volume but at much weaker amplitude, and the collective burst combines focal confinement with the strongest deposition. The collective deposition is the most confined, with axial FWHM of $11$, $13$, and $27$~\si{\micro m} for the collective, MPA, and linear regimes, respectively.

\begin{table}[tp]
\caption{Material constants (top) and excitation parameters (bottom) used for c-Si nanoparticles at $\lambda=1700$~nm.}
\label{tab:params}
\resizebox{\columnwidth}{!}{%
\begin{tabular}{lccl}
\toprule
Parameter & Symbol & Value & Unit \\
\midrule
\multicolumn{4}{l}{\textit{Material (c-Si)}} \\
TPA coeff.\textsuperscript{a} & $\beta_m$ & 1.0 & cm/GW \\
FCA cross section\textsuperscript{b} & $\sigma$ & 1.28 & $10^{-17}$~cm$^{2}$ \\
Auger coeff.\textsuperscript{c} & $\gamma$ & 3.8 & $10^{-31}$~cm$^{6}$/s \\
Critical carrier density\textsuperscript{d} & $n_\text{cr}$ & $10^{21}$ & cm$^{-3}$ \\
Damage threshold\textsuperscript{e} & $F_\text{th}$ & 0.2 & J/cm$^{2}$ \\
\midrule
\multicolumn{4}{l}{\textit{Excitation}} \\
Wavelength\textsuperscript{f} & $\lambda$ & 1700 & nm \\
Pulse duration & $\tau_\text{p}$ & 100 & fs \\
Burst duration & $\tau_\text{b}$ & $0.1\text{--}5$ & ns \\
Inter-pulse period & $\tau_\text{r}$ & $10\text{--}50$ & ps \\
Per-pulse fluence & $F_\text{p}$ & $0.02\text{--}0.1\,F_\text{th}$ & J/cm$^{2}$ \\
\midrule
\multicolumn{4}{l}{\textit{Nanoparticle geometry}} \\
Diameter & $d$ & 100 & nm \\
Surface recombination time\textsuperscript{g} & $\tau_\text{s}$ & 170 & ps \\
\midrule
\multicolumn{4}{l}{\textit{Thermal and acoustic}} \\
Si volumetric heat capacity\textsuperscript{h} & $C_\text{Si}$ & $1.66\times10^{6}$ & J/(m$^3$\,K) \\
Water thermal conductivity\textsuperscript{i} & $\kappa_\text{w}$ & 0.6 & W/(m\,K) \\
Water thermal diffusivity\textsuperscript{i} & $a_\text{w}$ & 0.143 & mm$^2$/s \\
Kapitza conductance\textsuperscript{j} & $G$ & 150 & MW/(m$^2$\,K) \\
Thermal relaxation time & $\tau_\text{th}$ & 2.5 & ns \\
Gr\"{u}neisen parameter (water)\textsuperscript{i} & $\Gamma_\text{w}$ & 0.12 & -- \\
Speed of sound (water)\textsuperscript{i} & $v_\text{s}$ & 1500 & m/s \\
Observation distance & $r$ & 100 & $\mu$m \\
\midrule
\multicolumn{4}{l}{\textit{Tissue and imaging (Monte Carlo)}} \\
Brain absorption\textsuperscript{k} & $\mu_\text{a}$ & 0.4 & mm$^{-1}$ \\
Brain scattering\textsuperscript{k} & $\mu_\text{s}$ & 2.3 & mm$^{-1}$ \\
Scattering anisotropy\textsuperscript{k} & $g$ & 0.87 & -- \\
Tissue refractive index & $n_\text{tis}$ & 1.37 & -- \\
Focal waist & $\omega_0$ & 2 & $\mu$m \\
Focal depth & $z_\text{f}$ & 500 & $\mu$m \\
Focal volume fraction & $f_\text{V}$ & $10^{-4}$ & -- \\
\bottomrule
\end{tabular}%
}
\par\smallskip
{\footnotesize\raggedright
\textsuperscript{a}Ref.~\cite{bristow2007}.\quad
\textsuperscript{b}Ref.~\cite{boggess1986}.\quad
\textsuperscript{c}Refs.~\cite{othonos1998, driel1986}.\quad
\textsuperscript{d}Ref.~\cite{pronko1998}.\quad
\textsuperscript{e}Ref.~\cite{bonse2002femtosecond-b0a}.\quad
\textsuperscript{f}Ref.~\cite{kobat-2009}.\quad
\textsuperscript{g}Ref.~\cite{grumstrup2014ultrafast-7ad}.\quad
\textsuperscript{h}Ref.~\cite{crchandbook}.\quad
\textsuperscript{i}Ref.~\cite{prost2015}.\quad
\textsuperscript{j}Refs.~\cite{gecahill2006, ramosalvarado2017}.\quad
\textsuperscript{k}Refs.~\cite{horton2013, Jacques2013}.\par}
\end{table}

\section{Application to Photoacoustic Imaging}

Finally, we consider the application of the collective energy deposition to photoacoustic imaging. Photoacoustic generation proceeds through an energy coupling chain in which the absorbed optical energy is converted to heat, the heat drives a thermoelastic expansion of the surrounding aqueous medium, and this transient expansion radiates an acoustic pressure wave~\cite{wang-2007}. We analyze the thermoelastic response from two perspectives, first the microscopic response of a single nanoparticle as the elementary source, then the macroscopic response of the nanoparticle ensemble within the focal volume. The absorbed energy density, $U$, deposited through collective absorption (Sec.~\ref{sec:collective_energy}) converts to lattice heat rapidly. The carriers' kinetic energy is transferred within picoseconds via carrier--phonon coupling, the stored band-gap energy follows upon nonradiative recombination within $\tau_\text{eff}$, and both are fast compared with the thermal relaxation time $\tau_\text{th}$ defined below~\cite{othonos1998,driel1986,hu2002}. A single nanoparticle thus acts as a nanoscale heat source driven by the burst, with $H \approx U$. Its temperature is spatially uniform ($\kappa_\text{w}/\kappa_\text{Si} \approx 0.004$), and Newton's law of cooling governs heat exchange with water~\cite{hu2002,prost2015}
\begin{equation}
    \frac{dT_\text{NP}}{dt} = \frac{1}{C_\text{Si}}\,\frac{d H}{d t} - \frac{T_\text{NP}}{\tau_\text{th}},
    \label{eq:newton}
\end{equation}
where $T_\text{NP}$ is the nanoparticle temperature rise above ambient, $d H/d t$ is the local heat power density inside the nanoparticle, equal to the absorbed power $d U/d t$ of Eq.~\eqref{eq:heating_continuous}, written with total derivatives because the spatially uniform particle renders $U$ and $H$ functions of time alone, $C_\text{Si} = \rho_\text{Si}c_\text{Si}$ is the volumetric heat capacity of silicon, with $\rho_\text{Si}$ the mass density and $c_\text{Si}$ the specific heat of silicon, and $\tau_\text{th}$ is the NP thermal relaxation time combining water-side diffusion~\cite{baffou2011} with the Si--water Kapitza resistance for interface conductance $G \approx \SI{150}{MW/(m^2\,K)}$~\cite{gecahill2006,ramosalvarado2017},
\begin{equation}
    \tau_\text{th} = \frac{C_\text{Si}\,d}{6}\left(\frac{d}{2\kappa_\text{w}} + \frac{1}{G}\right) \approx \SI{2.5}{ns}.
    \label{eq:tau_th}
\end{equation}
$T_\text{NP}$ is obtained by integrating Eq.~\eqref{eq:newton} numerically over the burst, with $d U/d t$ from Eq.~\eqref{eq:heating_modified} active during each pulse and $n_{\text{c},k}$ evolving via Eqs.~\eqref{eq:recurrence}. Unlike $n_\text{ss}$, which saturates through Auger-mediated negative feedback, $T_\text{NP}$ rises monotonically as successive pulses deposit energy faster than it can diffuse away on the timescale $\tau_\text{th}$.

The heat flowing out of the nanoparticle penetrates only a short distance into the surrounding water. Over the burst duration, thermal diffusion carries it a distance $\delta \approx \sqrt{a_\text{w}\,\tau_\text{b}} \approx 12$ nm (where $a_\text{w}$ is the thermal diffusivity of water), and even over the longer cooling tail $\tau_\text{th}$ this depth stays well below the particle size. The deposited heat is therefore confined to a nanometric shell of water around the particle, defined as the generating layer~\cite{prost2015}. As this layer heats up, it expands thermoelastically, and this expansion is what launches the pressure wave. The emission is dominated by the expansion of the water shell rather than of the nanoparticle itself~\cite{prost2015,wang-2008}, as confirmed experimentally by the near-vanishing of the photoacoustic signal at $4\,^\circ\text{C}$, where the thermal expansion coefficient of water vanishes~\cite{fukasawa2014,shinto2013}. The nanoparticle's own mechanical resonance, $f \sim v_\text{Si}/\pi d \approx 27$~GHz (with $v_\text{Si}$ the longitudinal sound speed in silicon), lies far above the bandwidth of the thermally generated signal (set by the nanosecond burst) and of any photoacoustic detection band, and such high-frequency vibrations are strongly damped in water. The particle's direct emission therefore does not contribute to the detected far-field signal~\cite{prost2015}. With $\delta \ll d \ll \lambda_\text{ac}$ (where $\lambda_\text{ac} \approx \tau_\text{b}\,v_\text{s}$), the heated shell radiates as an acoustic point absorber~\cite{calasso2001,prost2015}.  The nanoparticle discharges its stored heat into the water at the rate $C_\text{Si}V_\text{NP}\,T_\text{NP}/\tau_\text{th}$, the loss term of Eq.~\eqref{eq:newton} scaled by the nanoparticle volume $V_\text{NP} = \pi d^3/6$, and the far-field photoacoustic pressure of a single nanoparticle, $p_\text{NP}$, follows from the Calasso solution~\cite{calasso2001},
\begin{equation}
    p_\text{NP}(\mathbf{r},t) = \frac{\Gamma_\text{w}\,C_\text{Si}\,V_\text{NP}}{4\pi v_\text{s}^2\, r\,\tau_\text{th}}\,\frac{d T_\text{NP}}{d t},
    \label{eq:pa_pressure}
\end{equation}
where $\Gamma_\text{w} = \alpha_\text{th}\,v_\text{s}^2/c_{p,\text{w}}$ is the Grüneisen parameter of water at ambient temperature, $\alpha_\text{th}$ is the isobaric thermal expansion coefficient of water and $c_{p,\text{w}}$ is the specific heat capacity of water, and the right-hand side is evaluated at the retarded time $t - r/v_\text{s}$~\cite{calasso2001}. In this description, the nanoparticle acts as a heat reservoir that bridges two widely separated timescales, decoupling the femtosecond excitation from the nanosecond acoustic emission. The femtosecond pulses charge the reservoir almost instantaneously, whereas it discharges into the water only on the nanosecond scale set by the burst duration and the thermal relaxation time $\tau_\text{th}$. The water therefore responds not to the individual femtosecond pulses but to the slowly varying discharge of the stored heat, which follows $T_\text{NP}(t)$ and places the emission safely in the slow-heating limit of the point-source solution, valid when the heating time exceeds the thermal-acoustic time of water, $a_\text{w}/v_\text{s}^2 \approx \SI{60}{fs}$~\cite{calasso2001}. The residual per-pulse structure of $T_\text{NP}$ contributes only components above $1/\tau_\text{r} \gtrsim \SI{20}{GHz}$, which lie far outside any photoacoustic detection band and are strongly absorbed in water within micrometers, like the particle resonance discussed above. Equations~\eqref{eq:newton}--\eqref{eq:pa_pressure} thus amount to a deliberately simple approximation that combines the detailed treatments of Refs.~\cite{calasso2001} and~\cite{prost2015}, with the former providing the point-source acoustics and the latter the nanoparticle-to-water thermal coupling.

\begin{figure*}
    \centering
    \includegraphics[width=\linewidth]{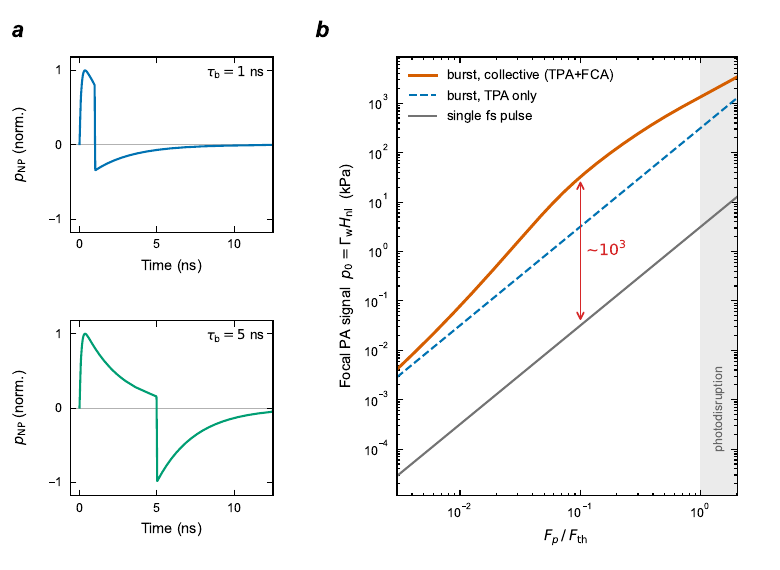}
    \caption{(a) Far-field photoacoustic waveform of a single nanoparticle for burst durations $\tau_\text{b}=1$ and $5$ ns at $\tau_\text{r}=10$ ps and $F_\text{p}=0.05\,F_\text{th}$, each waveform normalized to its peak. The waveform is bipolar in both cases, with the shape set by $\tau_\text{b}$ relative to the thermal relaxation time $\tau_\text{th}$: for $\tau_\text{b}<\tau_\text{th}$ the burst terminates while the positive lobe is still rising, leaving a shallow rarefaction, whereas for $\tau_\text{b}>\tau_\text{th}$ the positive lobe decays mid-burst and a near-equal negative transient follows at termination. The normalized shape is independent of $\tau_\text{r}$, which sets only the amplitude through $n_\text{ss}$ [Fig.~\ref{fig:dynamics}]. (b) Focal photoacoustic signal
$p_0 = \Gamma_\text{w}\,H_\text{nl}$ versus per-pulse fluence $F_\text{p}/F_\text{th}$ at matched per-pulse energy, for a single femtosecond pulse (TPA only), an $N$-pulse burst neglecting carrier accumulation (TPA only), and the collective interaction of the burst (TPA and accumulating FCA). The collective signal exceeds the single pulse by about $ N(1+\xi)$.}
    \label{fig:photoacoustic}
\end{figure*}

The chain from deposition to emission can now be closed in a single expression. Integrating Eq.~\eqref{eq:newton} gives $T_\text{NP}(t) = C_\text{Si}^{-1}\int_{-\infty}^{t}e^{-(t-t')/\tau_\text{th}}\,\dot U(t')\,dt'$, and substituting into Eq.~\eqref{eq:pa_pressure} yields
\begin{multline}
    p_\text{NP}(\mathbf{r},t) = \frac{\Gamma_\text{w}\,V_\text{NP}}{4\pi v_\text{s}^2\, r\,\tau_\text{th}}\biggl[\dot U(t) \\
    - \frac{1}{\tau_\text{th}}\int_{-\infty}^{t} e^{-(t-t')/\tau_\text{th}}\,\dot U(t')\,dt'\biggr],
    \label{eq:pa_from_U}
\end{multline}
again evaluated at the retarded time. After every intermediate quantity has dropped out, the far-field pressure is a direct image of the deposition rate. Thus the simple collective form of Eqs.~\eqref{nonlinear_to_collective_nonlinear} and~\eqref{eq:collective_scaling} is what the photoacoustic signal ultimately reports. The pressure waveform of Fig.~\ref{fig:photoacoustic} follows $\dot U$ while the burst runs and swings negative when the deposition stops at burst termination. Since the burst heats the nanoparticle cumulatively, the peak temperature limits the usable fluence, which must stay below the melting point of silicon. The ensemble of fluence values adopted throughout (Figs.~\ref{fig:dynamics}--\ref{fig:photoacoustic}) remains below this bound. In modeling the emission, we hold $\Gamma_\text{w}$ at its ambient value and neglect its temperature dependence for simplicity. However, as the temperature of the generating layer increases, the thermal expansion coefficient of water increases, raising $\Gamma_\text{w}$ and causing the emission to grow superlinearly with fluence~\cite{prost2015}. Driven hotter still, the generating layer can even nucleate transient vapor nanobubbles, whose expansion can further amplify the acoustic output well beyond thermal expansion alone~\cite{zharov2011nanobubble}.

Having obtained the single-particle response, the macroscopic photoacoustic response of the nanoparticle suspension follows the standard treatment of laser-irradiated colloidal suspensions, in which the focal volume containing many nanoparticles is described as an effective bulk photoacoustic medium under stress confinement at the focal scale, $\tau_\text{b} \lesssim \omega_0/v_\text{s}$~\cite{shinto2013,fukasawa2014,wang-2008}. This condition holds for $\omega_0 = \SI{2}{\micro m}$ and $\tau_\text{b} = 1$ ns. We characterize the nanoparticle loading by the focal volume fraction $f_\text{V}$, the fraction of the focal region occupied by silicon nanoparticles. The nonlinear (TPA and FCA) contributions to the deposited heat scale linearly with $f_\text{V}$, whereas the linear tissue absorption is independent of it.

Within the focal volume under stress confinement, the local thermoelastic pressure rise follows the classical photoacoustic relation~\cite{wang-2008,wang-2007}
\begin{equation}
    p_0(\mathbf{r}) = \Gamma_\text{w}\,H(\mathbf{r}),
    \label{eq:p0_local}
\end{equation}
with $H(\mathbf{r}) = U(\mathbf{r})$ the absorbed energy density of Fig.~\ref{fig:confinement}. This energy comprises a delocalized linear tissue background and a confined nonlinear (TPA and FCA) contribution $H_\text{nl}$. The corresponding nonlinear focal signal $p_0 = \Gamma_\text{w}\,H_\text{nl}$ is the quantity plotted in Fig.~\ref{fig:photoacoustic}(b). The acoustic detection and image reconstruction that follow form the second stage of the photoacoustic forward problem and are treated separately in the standard framework~\cite{wang2016practical}. Separating $H_\text{nl}$, a confined, time-gated, intensity-nonlinear signal, from the delocalized linear tissue background is an experimental consideration rather than part of the enhancement mechanism studied here. It can be accomplished with established techniques~\cite{lai2014nonlinear,yao2014photoimprint,danielli2014labelfree,kaneko2025signal}, on which we rely rather than developing new ones.

Figure~\ref{fig:photoacoustic}(b) quantifies the enhancement of the focal photoacoustic signal at matched per-pulse energy. We track the enhancement up to the operating fluence $F_\text{p}=0.10\,F_\text{th}$, the largest per-pulse value kept below the melting bound, while the spatial and temporal profiles of Figs.~\ref{fig:confinement} and~\ref{fig:photoacoustic}(a) use an illustrative $F_\text{p}=0.05\,F_\text{th}$, whose shapes are essentially fluence-independent. Tolerable loadings depend on the specific silicon formulation and application~\cite{park2009}, so we adopt a representative focal volume fraction $f_\text{V}=10^{-4}$ ($\sim0.2$~\si{mg/mL} silicon). The absolute signal scales linearly with $f_\text{V}$, and at this loading the focal volume holds only a modest number of nanoparticles, so $p_0 = \Gamma_\text{w}\,H_\text{nl}$ describes the focal source coarse-grained over the particle distribution, with the discreteness contributing shot-to-shot variance rather than changing the mean signal. The comparison itself is entirely within nonlinear excitation. A single femtosecond pulse and the collective burst are built from identical pulses, the burst differing only in firing $N$ of them at GHz spacing. Relative to a single pulse, the burst's focal signal is larger by a factor $\sim N(1+\xi)$, which factors into a trivial but crucial factor of $N$ from the pulse count and the genuine collective gain factor of $(1+\xi)\sim 9$ at this operating fluence, the result of the accumulated free-carrier absorption, the difference between the GHz burst and the same pulses fired without carrier accumulation. This collective gain is set by the FCA-to-MPA ratio $\xi$ mapped in Fig.~\ref{fig:dynamics}(b). It is independent of the nanoparticle loading, and retains the focal confinement of nonlinear excitation. For the $N=100$ bursts considered here, the two factors together raise the focal signal by nearly three orders of magnitude over a single pulse of the same per-pulse energy.

\section{Discussion}

The collective nonlinear absorption mechanism reported here combines the benefits of the tighter spatial confinement of multiphoton absorption and the greater ease of energy coupling of linear absorption. Early pulses in a burst seed a feedback-driven buildup of a non-equilibrium population of free carriers via multiphoton excitation. This opens a burst-gated transient window in which the dominant channel becomes effectively linear absorption by those carriers, which remain confined to the multiphoton-defined focal volume. Each pulse arriving before the carriers recombine adds to a non-equilibrium population that the burst creates and sustains and that dissipates and returns to equilibrium once the burst ends~\cite{radu2011transient, krishnamoorthy2019optical, fang2017direct}, making the enhancement intrinsically transient. The carrier density approaches a steady state fixed by the balance between per-pulse injection and recombination, and because it is reached only where multiphoton absorption has seeded it, the efficient linear channel inherits the spatial selectivity of the nonlinear one.

Coupling multiphoton carrier generation with Auger and surface recombination, the analytical model yields a piecewise-exact description of the carrier dynamics and condenses the entire burst enhancement into a single compact factor, $N(1+\xi)$. Here $N$ is the number of pulses and $\xi$ measures how strongly the accumulated free-carrier absorption dominates the seeding multiphoton channel, with $N \sim 10^2\text{--}10^3$ and $\xi \sim 10$ under the conditions considered here. Taken at matched per-pulse energy, this factor reflects the carrier dynamics alone rather than the way the deposited energy is later turned into a signal, so the same collective gain is expected whatever the target modality. Reaching the collective regime requires only that the burst deliver GHz-rate pulses, so that the inter-pulse spacing falls below the carrier recombination time, while each pulse stays below the single-pulse damage threshold, conditions already within reach of existing burst lasers~\cite{kerse2016} and of well-characterized silicon nanoparticles~\cite{ofarrell2006, peng2014}. 

Viewed this way, collectivity converts an $m$-photon excitation into an effectively $(m+1)$-photon one, an extra order of nonlinearity gained at virtually no cost in cross-section, which in the long run may be the most consequential aspect of the mechanism. More generally, the increase by one in nonlinear order arises only because the nonlinear absorption gates a linear absorption process. If instead an $m$-photon process creates a transient state that enables a subsequent $n$-photon interaction, Eq.~\ref{nonlinear_to_collective_nonlinear} becomes $I^n(t)\int_{-\infty}^t I^m(t')dt'$, resulting in effectively a $(m+n)$-photon process. 

Although we developed a detailed application example for photoacoustic imaging, the collective absorption mechanism is not limited to this application. In multiphoton fluorescence microscopy~\cite{Denk-1990, Helmchen-2005} it could raise excitation efficiency without sacrificing optical sectioning~\cite{centonze1998multiphoton-904}, provided the nanoparticles are paired with a thermoresponsive fluorescent reporter that converts the burst-driven heating into emission~\cite{brites2012, jaque2012, zhou2020, okabe2012}, and the same particles would let photoacoustic and fluorescence contrast be recorded together~\cite{li2017, chen2024}. Photothermal therapy~\cite{Huang2006} is another natural target, where collective heating could reach therapeutic temperatures at reduced peak intensity while keeping the deposited energy confined. Looking further, the same idea of using early pulses to open a transient channel for later ones may carry over to other nonlinear processes, from THz generation~\cite{castrocamus2005, lu2022} to a self-sustained index modulation that guides successive pulses of the burst through a transiently written waveguide~\cite{couairon2007, chen2012solitons, majus2014}. In each case the underlying conceptual change is the same, from a burst treated as a train of independent pulses to one in which the pulses act collectively, which is what allows efficiency and spatial confinement to be gained together rather than traded against each other.

\begin{acknowledgments}
This work was supported by the Alexander von Humboldt Foundation through an Alexander von Humboldt Professorship awarded to F. Ö. Ilday and the European Research Council (ERC) under the European Union’s Horizon Europe research and innovation programme (grant agreement no. 101055055, ERC Advanced Grant UNILASE).
\end{acknowledgments}

\bibliography{references}

\end{document}